\documentstyle[aps,epsf,twocolumn,prl]{revtex}
\def\beqn       {\begin{eqnarray}}
\def\eeqn       {\end{eqnarray}}
\def\bmat       {\left( \begin{array}}
\def\emat       {\end{array} \right)}

\def\f		{\frac}

\def\g		{\gamma}
\def\l		{\lambda}
\def\s		{\sigma}
\def\th		{\theta}

\def\bA		{{\bf A}}
\def\bB		{{\bf B}}
\def\bR		{{\bf R}}

\def\vs         {{\mbox{\boldmath $\s$}}}

\def\sss	{\scriptscriptstyle}
\def\nR		{\nabla_{\!\sss R}}

\def\ciu	{c_{i\uparrow}^{}}
\def\ciud	{c_{i\uparrow}^\dagger}
\def\cid	{c_{i\downarrow}^{}}
\def\cidd	{c_{i\downarrow}^\dagger}
\def\Sz		{S_z(i)}
\def\Sp		{S_+(i)}
\def\Sm		{S_-(i)}

\def\lp		{\grave{}}
\def\rp		{\acute{}}

\newcommand{\bra}[1]	{\langle #1|}
\newcommand{\ket}[1]	{|#1\rangle}
\newcommand{\av}[1]	{\langle #1\rangle}

\begin{document}
\draft
\twocolumn[\hsize\textwidth\columnwidth\hsize\csname @twocolumnfalse\endcsname
\title{Berry Phases with Real Hamiltonians With and Without a 
Many-body System as a Background}
\author{S. P. Hong, H. Doh, and S. H. Suck Salk}
\address{Department of Physics\\
  Pohang University of Science and Technology,
  Pohang 790-784, Korea}
\maketitle
\begin{abstract}
We present both the gauge theoretic description
and the numerical calculations of the Berry phases
with the real eigenstates, involving one with a many-body system
as a background and the other with no such background.
We demonstrate that for the former
the sign of the Berry phase factor for a spin $\f{1}{2}$ particle (hole) 
coupled to a slow subsystem (phonon)
depends on both the strength of electron correlations
and the characteristics of the closed paths,
unlike the cases for the latter.
\end{abstract}
]

The real wave functions in association with real Hamiltonians
are frequently encountered in numerous physical problems,
not to speak of condensed matter physics.
There exists still an unresolved problem in defining
the Berry phase concerned with the real eigenfunctions
in that the gauge potential cannot be defined.
The Berry phase \cite{Ber,Sim,KI,MT} arises from a $U(1)$ gauge potential 
as a result of the adiabatic transition
involving the single-valued, complex, non-degenerate eigenstates
of a fast subsystem coupled to a slow subsystem.
In the following we briefly draw an attention to the main issue
of the present work.

In a system described by the time-dependent Hamiltonian $H(\bR(t))$
through a slowly varying parameter $\bR$,
\beqn
H(\bR(t)) \ket{n;\bR(t)} = E_n(\bR(t)) \ket{n;\bR(t)} ~,
\label{eqmotion}
\eeqn
the total phase change of the state $\ket{\psi}$ round a closed
loop for a time period of $T$ is given by
\beqn
\ket{\psi(T)} = \exp[i\g_n(C)]
  \exp\left\{-\f{i}{\hbar}\int_0^T dt E_n(\bR(t))\right\} \ket{\psi(0)} ~,
  \nonumber
\eeqn
where the Berry phase $\g_n(C)$ in the first factor is given by
\beqn
\g_n(C) = i\oint\bra{n;\bR}\nR\ket{n;\bR}\cdot d\bR ~,
\label{bergam}
\eeqn
and 
$\ket{\psi(0)}=\ket{n;\bR(0)}$ at $t=0$.
To obtain a non-zero value of $\g_n$, the above eigenstate $\ket{n;\bR(t)}$
needs to be single-valued, complex and non-degenerate.
Thus one cannot define the gauge potential
\beqn
\bA(\bR) = i\bra{n;\bR}\nR\ket{n;\bR}
\eeqn
from the use of the real eigenfunctions.

In order to allow for the case of multi-valuedness
of the complex eigenstates $\ket{n;\bR}$
Berry\cite{Ber} derived the following expression of the `magnetic field' 
from (\ref{bergam}),
\beqn
\lefteqn{\bB(\bR) = } \nonumber \\
&&  -\mbox{Im}\sum_{m\neq n}
  \f{\bra{n;\bR}(\nR H)\ket{m;\bR}\times\bra{m;\bR}(\nR H)\ket{n;\bR}}
  {[E_m(\bR)-E_n(\bR)]^2}~.
\label{magnetic} 
\eeqn
To avoid a confusion found in the literature,
once again we stress that $\ket{n;\bR}$ above is the complex (but not real) 
and multi-valued eigenstate.
From this expression one finds the existence of `magnetic monopole(s)'
corresponding to the degenerate eigenstates
$\ket{m;\bR^{*}}$ at singular point(s), 
$\bR=\bR^*$ in the parameter space $\bR$.

The objective of the present study is two-fold; one is to rigorously discuss, 
in terms of the gauge potential, 
the Berry phase with the multi-valued real eigenfunctions,
and the other, to newly examine the variation of the Berry phase 
with the strength of electron correlations 
for a coupled system made of 
a hole, the fast subsystem, and a phonon, the slow subsystem.
In addition, we discuss the two distinctively different cases
from which the Berry phase arises;
one from a many-body background system 
and the other from no such background.

As mentioned above, the magnetic field $\bB$ is not definable
with the use of the multi-valued real eigenstates $\ket{n;\bR(t)}$,
since the gauge potential $\bA(\bR)$ vanishes.
However, by a proper gauge transformation it is possible
to define the Berry phase as will be discussed below. 
The gauge invariance of the equation of motion (\ref{eqmotion}) 
necessitates the local gauge transformation of the form
\beqn
\ket{n;\bR}\rp = e^{i\l_n(\bR)}\ket{n;\bR}\label{1_1}
\label{gauget}
\eeqn
The Berry phase $\g_n(C)$ is then
\beqn
\g_n &=& i\oint \lp\bra{n;\bR}\nR\ket{n;\bR}\rp \cdot d\bR
  \nonumber \\
  &=& -\int^{\bR(T)}_{\bR(0)} d\l_n(\bR) = -[\l_n(\bR(T)) - \l_n(\bR(0))] ~.
\label{gamma}
\eeqn

We now compute the magnetic field ${\bf B}(\bR)$,
\beqn
\bB(\bR) &=& \nR \times \bA
  = \nR\times i\;\lp\bra{n;\bR}\nR\ket{n;\bR}\rp \nonumber \\
  &=& -\mbox{Im}\sum_{m\neq n}
  \lp\av{\nR n;\bR|m;\bR}\rp \times \lp\bra{m;\bR}\nR\ket{n;\bR}\rp
\label{newB}
\eeqn
with $m\neq n$ since 
$\lp\av{\nR n;\bR|n;\bR}\rp\times \lp\bra{n;\bR}\nR\ket{n;\bR}\rp$
is a real vector.
We find from (\ref{eqmotion})
\beqn
\lp\bra{m;\bR}\nR\ket{n;\bR}\rp
= \f{\lp\bra{m;\bR}(\nR H)\ket{n;\bR}\rp}{E_n-E_m} ~, ~~~m\neq n ~.
\label{mdeln}
\eeqn
The insertion of (\ref{mdeln}) into (\ref{newB}) 
and a refinement of its result leads to
\beqn
\lefteqn{\bB(\bR) =} \nonumber \\
&&   -\mbox{Im}\sum_{m\neq n}
  \f{\bra{n;\bR}(\nR H)\ket{m;\bR}\times\bra{m;\bR}(\nR H)\ket{n;\bR}}
  {[E_m(\bR)-E_n(\bR)]^2}~.
\label{Breal}
\eeqn
Here both $\ket{m;\bR}$ and $\ket{n;\bR}$ are now the real
eigenstates unlike the ones in Expression (\ref{magnetic}).
It is important to realize from (\ref{newB}) and (\ref{Breal}) 
that ${\bf A}(\bR)\neq 0$ but ${\bf B}(\bR)=0$ 
at other than the degenerate (singular) point(s) $\bR = \bR^*$.
Thus the Berry phase for the real eigenstates is realized
as a type of the Aharonov-Bohm phase.

Let us first examine a one-body problem involving no background. 
For a particle with spin in the constant magnitude of 
a magnetic field with slowly varying orientation constrained  
in the 2-D plane, the Hamiltonian becomes real;
$H={\vs}\cdot \bR=\bmat{cc} R_3& R_1\\ R_1 &-R_3\emat$ 
with $\vs$, the Pauli spin matrix. 
Choosing $\l_n(\theta)=nk\theta$ with $k$ being a non-zero integer 
we write the gauge transformation,
\beqn
\ket {n;\theta}\rp=\exp(ink\theta)|n ;\theta\rangle ~.
\label{gt}
\eeqn
The insertion of (\ref{gt}) into (\ref{gamma}) leads to 
the Berry phase of $\g_n=-2\pi n k$.
The eigenstate of an electron coupled to
the magnetic field $\bB$ confined in the 2-D $x$-$z$ plane
is real and double-valued.
For the double-valued real eigenstates of 
$\ket{n;\th}=\bmat{c} \cos\th/2\\ \sin\th/2 \emat$
with $n=\f{1}{2}$ as a spin component along the magnetic field, 
the Berry phase factor round the closed path which encircles
$k$ (many) `solenoids' or  a single solenoid of strength $k$
is then $\g_n(C)=-k\pi$.
The Berry phase is thus equivalent to  the flux 
in the `magnetic solenoid' with `strength' $-n$ 
which vertically pierces through the 2-D plane
if there exists only one singular point in the plane.
In the present case 
of a particle with no background,
the Berry phase is independent of the choice of a path
as long as the closed path encircles the solenoids.
For the single unit of the solenoid, i.e., $k=1$, 
the Berry phase factor is $\g_n(C)=-\pi$ as expected.
This is simply an Aharonov-Bohm phase type 
through the closed path 
which encloses the single solenoid $k=1$.
It is now realized that the geometric phase (Berry phase)
with the real eigenstates is composite in nature 
as it is determined by both the intrinsic property,
i.e., the spin of the particle 
and the magnetic flux of the solenoid.
Later we will examine the intrinsic property (spin) of a particle (hole)
coupled to a phonon in the many-body background 
of antiferromagnetic spin correlations.
It is now clear that for the real eigenstates $\ket{n;\bR}$
the local gauge transformation 
$\ket{n;\bR}\rp = e^{i\l_n(\bR)}\ket{n;\bR}$
is essential for realizing the presence of the gauge potential and  
the `magnetic solenoid'
as in the case of the Aharonov-Bohm phase.

Let $S$ be a simply connected surface bounded by a closed loop
 in a parameter space $\bR$.
If the phase change of the eigenstates $\ket{n;\bR}$
occurs when transported adiabatically round a closed loop on $S$ 
there must be at least one singular point
where $\ket{n;\bR}$ is discontinuous,
due to the intersection of potential energy surfaces \cite{MT,HL,Sto}.
In the following we show an interesting case of 
the Aharonov-Bohm type Berry phase with the choice of 
a many-body system as a background,
in which there can be any number $k$
of singular points through which the `magnetic solenoids'
pass vertically through the 2-D plane.

By considering the Holstein-type electron (hole)-phonon coupling
\cite{Hol,EH,Hub}
we use the effective single-band
Holstein-Hubbard model Hamiltonian \cite{ZS}
for a two-dimensional square lattice,
\beqn
H(\bR) &=& -t\sum_{<ij>,\s}c_{i\s}^\dagger c_{j\s}
  + U\sum_i n_{i\uparrow} n_{i\downarrow}
  \nonumber \\
&~&  - g\sum_i n_i R_i 
  + \f{K}{2}\sum_i\{(\Delta_i^x)^2+(\Delta_i^y)^2\}\label{1_2}
\eeqn
with $n_{i\s} = c_{i\s}^\dagger c_{i\s}$.
Here the real Hamiltonian $H(\bR)$ depends on the time-dependent
external parameter $\bR$ corresponding to local lattice distortions.
$c_{i\s}^\dagger ~(c_{i\s})$ is the creation (annihilation)
operator of an electron with spin $\s$ at site $i$.
$t$ represents the electron hopping strength;
$U$, the electron-correlation strength;
$g$, the electron (hole)-phonon coupling constant
and $K$, the spring constant.
The local Holstein distortion $R_i$ for an in-plane
breathing mode 
at the lattice site $i=(i_x,i_y)$ is defined by \cite{RFB,FRMB}
\beqn
R_i = \Delta_{i}^x - \Delta_{i-(1,0)}^x
  + \Delta_{i}^y - \Delta_{i-(0,1)}^y ~,
\eeqn
where $\Delta_{i}^x ~ (\Delta_{i}^y)$ is the displacement
of oxygen from equilibrium along the positive $x$ ($y$)
direction at the lattice site $i$ in the CuO$_2$ plane and the other two, 
along the negative $x$ $(y)$ direction as shown in Fig.~1.
Unlike the Holstein-$tJ$ model Hamiltonian \cite{RFB,FRMB,SYZ}
the Holstein-Hubbard model Hamiltonian \cite{ZS}
above has an advantage of investigating
the geometric phase (Berry phase)
varying with the strength of correlation in the background.

\begin{figure}[hbt]
\epsfxsize=6.0cm
\epsfysize=6.0cm
\centerline{\epsffile{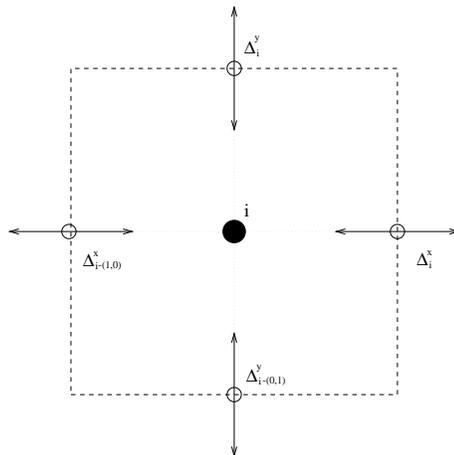}}
\caption{CuO$_4$ unit in the CuO$_2$ planar square lattice
  to show a local lattice distortion (breathing mode).
  The solid circle denotes copper and the open circle, oxygen.}
\end{figure}

To obtain the Berry phase for the real 
ground state wave function,
we treat a $4\times 4$ periodic lattice with a hole 
surrounded by the background of nearly half-filled antiferromagnetic spins.
To describe the propagation of the local lattice distortions 
round a closed path, we write 
\beqn
R_i(\tau) = R_{{\sss A},i} + (R_{{\sss B},i}-R_{{\sss A},i}) \tau
\eeqn
with $\tau$ being the dimensionless time lapse, $0\le\tau\le 1$
between the lattice sites; the lattice distortion (breathing) occurs at
site $A$ at $\tau=0$ and at site $B$ at $\tau=1$.
In this manner, a closed path in the parameter space $\bR$ can
be defined.

The mean field (Hartree-Fock) approximation \cite{Cho} of (11) is  
\beqn
H &=& -g\sum_i \psi_i^\dagger\psi_i R_i(\tau) -t\sum_{<ij>}\psi_i^\dagger \psi_j
  \nonumber \\
&~&  + U\sum_i \psi_i^\dagger[\f{1}{2}\av{n(i)} - \av{S(i)}]\psi_i
  \nonumber \\
&~& +U\sum_i[\av{\Sz}^2+\av{\Sp}\av{\Sm}-\f{1}{4}\av{n(i)}^2] ~,
\label{linham}
\eeqn
where the two component spinor is 
$\psi_i = \bmat{c} \ciu \\ \cid \emat$
and the spin operators are defined by 
$\Sz = \f{1}{2} (\ciud\ciu - \cidd\cid)$,
$\Sp = \ciud\cid$,
$\Sm = \cidd\ciu$,
and
$n(i) = \ciud\ciu + \cidd\cid$.

By minimizing the total energy with respect to the average
spin densities $\av{\Sz}$, $\av{\Sp}$ and $\av{\Sm}$,
the linearized Hamiltonian (\ref{linham})
was self-consistently treated
for the lattice of the periodic $4\times 4$ unit cell.
For various values of $U$ and $g$ between $0t$ and $10t$ we evaluated 
the Berry phase factors.
Sch\"{u}ttler et~al. \cite{SYZ} employed the Holstein-$tJ$ model by choosing 
sufficiently strong hole-phonon coupling.
Our computed results of the Berry phase factor 
in the region of strong correlation 
and large hole-phonon coupling 
agreed with their results 
for the case of one-hole tunneling with various paths shown in Fig.~2.

For numerical evaluations of the Berry phase factor 
we used the path integral approach \cite{KI,SYZ};
the adiabatic quantum transition involving  the eigenstate $\ket{n;\bR}$
round a closed path is given by 
\beqn
T_{nn}(C) &=& \exp\left[-\f{i}{\hbar}\int_0^T E_n(\bR(t))dt\right]
  \av{n;\bR(T)|n;\bR(0)} ~,
  \nonumber \\
&&
\label{1_4}
\eeqn
where the first factor is the dynamic phase factor 
and the last one, the geometric phase (Berry phase) factor given by 
\beqn
\lefteqn{\av{n;\bR(T)|n;\bR(0)}\equiv} \nonumber \\
&& \lim_{N\rightarrow\infty} \sum_{k=1}^N
  \av{n;\bR(t_k)|n;\bR(t_{k-1})} = e^{i\g_n(C)} ~.
\label{1_3}
\eeqn

Only for the single-valued complex eigenstates $\ket{n;\bR}$
the Berry phase $\g_n$ can be obtained directly from the relations
(\ref{bergam}) and (\ref{magnetic}).
For the real eigenstate with the Hamiltonian (\ref{linham})
we evaluated the adiabatic Berry phase factor 
(\ref{1_3}) by computing the ground state electronic wave functions
at each time step $t_k$.
Lately we \cite{Sal} found that in scattering processes the role 
of non-adiabatic intermediate transitions  may be important. 
In such cases, the expression (\ref{1_3}) is invalid. 

\begin{figure}[hbt]
\epsfxsize=6.0cm
\epsfysize=6.0cm
\centerline{\epsffile{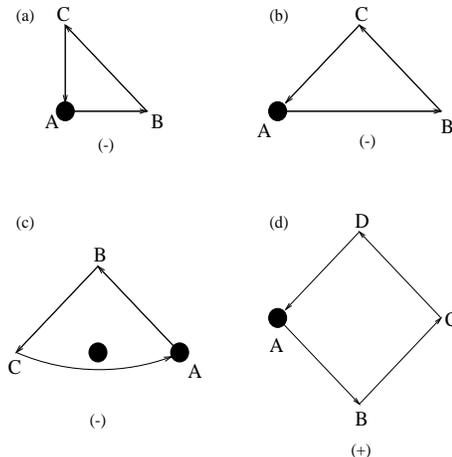}}
\caption{Hole hopping paths; the solid circle represents
  the initial location of the hole.
  Each sign denotes the Berry phase factor, $+1$ or $-1$.}
\end{figure}

In Fig.~1 a single CuO$_4$ unit \cite{RFB,FRMB} that appears in the CuO$_2$
planar square lattice of high $T_c$ copper-oxide superconductors 
is drawn to exhibit 
the local lattice distortions $\Delta_i^{x,y}$
involving a copper atom at the center
and its four surrounding oxygen atoms.
By choosing the strong correlation of $U=6t$ 
and the electron-phonon coupling of $g=6t$
for various closed paths,
the sign of the computed Berry phase factors 
are shown in Fig.~2.
In Fig.~3
we computed each single particle
state involving 7 electrons with spin-up
and 8 electrons with spin-down respectively 
for the doping case of one hole with spin-up 
for the path shown in Fig.~2(b).
Due to the limited space we show only the lowest unoccupied
energy level denoted as a dotted line for the spin-up electrons.
A hole coupled to the local lattice distortion (breathing mode) 
may move around with some possibility of making a closed loop
in the two-dimensional square lattice.
A midway point (denoted as $\uparrow$ in Fig.~3) is 
where a symmetric lattice distortion occurs
between two neighboring  sites, say, $A$ and $B$. 
At such midway points the energy gap 
between the HOMO (the highest occupied molecular orbital)
and the LUMO (the lowest unoccupied molecular orbital)
is found to be a minimum, 
indicating that there exist degenerate points 
in the lattice distortion parameter space $\{R_i\}$.
The HOMO-LUMO gap becomes a maximum at sites $A$, $B$ and $C$
which correspond to the symmetry breaking points. 

In the case of relatively strong electron correlations (large $U$),
the computed Berry phase factor
was $-1$ for all the triangular closed paths
and $+1$ for the closed path of the square shown in Fig.~2.
We found that there exists an odd number of degenerate points (singular points),
namely 3 of them for the former and an even number of degenerate
points, i.e., 4 for the latter.
For the case of sufficiently weak correlations, say, $U\le t$,
it is quite interesting to note 
that the computed Berry phase factor is $+1$ even for the
triangular closed paths,
contrary to the case of strong correlations. 
Thus our findings are as follows; 
the hole behaves as a particle of spin 0 
for the background of weak correlations 
and as a spin $\f{1}{2}$ fermion
for the background of relatively strong correlations.
We further found that the Berry phase factor for the hole 
which behaves as a spin $\f{1}{2}$ fermion 
in the background of strong correlations 
is subject to change its sign depending on the
nature of closed paths, 
i.e., the number of degenerate
points that can exist round the closed paths in the parameter space.

\begin{figure}[hbt]
\epsfxsize=8.0cm
\epsfysize=7.0cm
\centerline{\epsffile{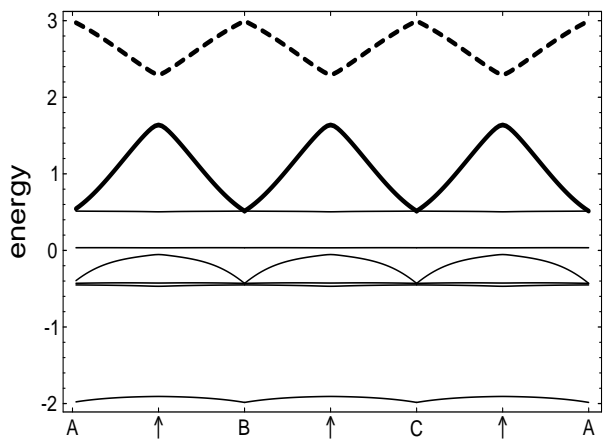}}
\caption{Energy levels corresponding to lattice distortion
  along the closed path ABCA in Fig.~2(b).
  The dotted line indicates the lowest unoccupied energy level (LUMO)
  and the thick solid line, the highest occupied energy level (HOMO).
  Midway points (denoted as $\uparrow$) represent symmetric lattice distortions
  between neighboring sites $i$ and $j$ ($i,j=A,B,C$).}
\end{figure}

The direct use of the multi-valued real eigenstates
for defining the `magnetic flux' or the Berry phase is not justifiable.
We argued, on the two grounds, the validity of the local
gauge transformation $\ket{n;\bR}\rp = e^{i\l_n(\bR)}\ket{n;\bR}$
with $\ket{n;\bR}$, the real eigenstate; 
namely the gauge invariance of the equation of motion (1)
and the realization of $\l_n$ as a 
`Aharonov-Bohm' phase type.
In short, using the local gauge transformation
we were able to define the `magnetic flux' of a solenoid 
and thus the Berry phase of the Aharonov-Bohm phase type.
We then explored
the Berry phase for the adiabatic transitions 
involving a hole coupled to the lattice distortion (breathing mode) 
in the antiferromagnetic background 
of both weakly and strongly correlated electrons.

The following findings are in order.
The strength of electron correlations affects 
the sign of the Berry phase factor;
round the simplest triangular closed paths
the Berry phase factor of the hole is $+1$
for the case of weak correlation,
indicating the hole state of spin 0, 
and $-1$ for the case of strong correlation, 
indicating the hole state of spin $\f{1}{2}$. 
However the Berry phase factor for the spin $\f{1}{2}$ fermionic hole
depends on the nature of the closed paths, e.g., 
$-1$ for the triangular closed paths
and $+1$ for the square closed path.
This is because the sign of the Berry phase factor 
is determined by the number $k$ of degenerate points (singular points) 
at which the fluxes pierce through the 2-D plane,
depending on the characteristics of closed paths 
in the parameter space.
Thus this is quite different from the system 
with no background 
in which the sign of the Berry phase does not depend on
the choice of closed paths, as was discussed earlier.

One (S.H.S.S) of the authors is highly grateful
to Professor Han-Yong Choi 
for rendering him an essential part of his current computer code.
He acknowledges the financial supports of
the Korean Ministry of Education BSRI (1996)
and POSTECH/BSRI special programs.
He is also grateful to the SRC program (1996) of 
the Center for Molecular Science at KAIST.

\end{document}